\documentclass[12pt,preprint]{aastex}

\def\eg{{e.g.\ }}

\def\ie{{i.e.\ }}
\def\etal{\mbox{\it et al.\,}}

\newcommand{\gsc}{\;\mathrm{g~cm^{-2}}}
\newcommand{\cm}{\; {\rm cm}}

\newcommand{\AU}{\; {\rm AU}}

\begin{document}

\title{Planetesimal Formation by Gravitational Instability}

\author{Andrew N. Youdin\altaffilmark{1}}
\affil{Department of Physics, 366 Leconte Hall, University of California,
Berkeley, CA 94720}

\author{Frank H. Shu\altaffilmark{1}}
\affil{Department of Physics, National Tsing Hua University,
101, Sec. 2, Kuang Fu Rd., Hsinchu 30013, Taiwan, R. O. C.} 

\altaffiltext{1}{Also: Department of Astronomy, 601 Campbell Hall,
University of California, Berkeley, CA 94720}
\email{youd,shu@astron.berkeley.edu}

\begin{abstract}
We investigate the formation of planetesimals via the gravitational
instability of solids that have settled to the midplane of a
circumstellar disk.  Vertical shear between the gas and a subdisk of
solids induces turbulent mixing which inhibits gravitational
instability.  Working in the limit of small, well-coupled particles,
we find that the mixing becomes ineffective when the surface density
ratio of solids to gas exceeds a critical value.  Solids in excess of this
precipitation limit can undergo midplane gravitational instability and
form planetesimals.  However, this saturation effect typically
requires increasing the local ratio of solid to gaseous surface
density by factors of two to ten times cosmic abundances, depending on
the exact properties of the gas disk.  We discuss existing astrophysical
mechanisms for augmenting the ratio of solids to gas in protoplanetary
disks by such factors, and investigate a particular process that
depends on the radial variations of orbital drift speeds 
induced by gas drag.  This
mechanism can concentrate millimeter sized chondrules to the
supercritical surface density in $\leq {\rm few}
\times10^6$ years, a suggestive timescale for the disappearance
of dusty disks in T Tauri stars.  We discuss the relevance of our results to
some outstanding puzzles in planet formation theory: the size of the
observed solar system, and the rapid type I migration of Earth mass
bodies.
\end{abstract}

\section{Introduction}   
\label{sec:intro}
The planetesimal hypothesis states that terrestrial planets and the
icy cores of gas giants formed in a disk by the accretion of smaller
solid bodies, called planetesimals.
Planetesimals are defined as primitive solids of kilometer
size or larger.  Such a scale is meaningful because it delimits
where pairwise gravity dominates gas drag,
leading to the runaway or oligarchic growth of protoplanets  \citep{lis93}.
Because planetesimal formation must occur, at least for
the giant planets, while gas is still present in the
solar nebula, the coupling of gas and solids may be crucial
to understanding the process.

The prevailing view for planetesimal formation posits agglomerative
growth from sub-micron-sized ``interstellar'' grains to km-sized
bodies (cf. the review of Lissauer 1993).  However, building
planetesimals through pure solid-state sticking forces has
many problems.

First, in the inner solar system, interior to the ``snow line,'' the
only solids available for the formation of the terrestrial planets
were rocks.  Everyday experience tells us that dry silicate
particulates of millimeter and larger size, such as sand, do not stick
at almost any speed of attempted assemblage.  Measurements of
interparticle collisions in microgravity experiments reinforce these
intuitive impressions; indeed, $\mu$m sized dust particles disrupt
larger aggregates upon collision at relative velocities $>$ m/s
\citep{bw00}.  For collisions between fluffy mm-sized agglomerations
of $\mu$m sized grains, disruption also occured at speeds $>$ m/s.
However at lower speeds, down to 15 cm/s, only restitution (bouncing)
was observed, without any sticking \citep{bm93}.  Relative velocities
considerably higher than 1 m/s arise from chaotic motions in a
turbulent disk or from differential orbital drift in a laminar disk
\citep{wc93}.  For mm-sized crystalline silicates, such as chondrules
(see below), significant shattering occurs only at collision speeds
$>$ several km/s \citep{jth96}.  Thus, compact pieces of rock that
make up the bulk of the material of chondritic meteorites, whose
parent bodies are primitive asteroids (rocky planetesimals), are
unlikely either to agglomerate or to fragment at the collisional
velocities common in the nebular disk when there is still appreciable
gas present to exert appreciable drag on the solids (the well-coupled
limit, see Appendix).

Second, at temperatures significantly below their melting points, even
ices in the outer solar system may not be much more sticky.
Experiments by \citet{sup97} found that water frost has spring-like
properties and can only induce sticking for collision speeds $< 0.5$
cm/s.  Furthermore, if ice balls could grow by continued agglomeration
until disruptive tidal forces became stronger than the cohesive
strength of crystalline ice, we should expect many icy particulates in
Saturn's rings to acquire sizes of order $\sim 10-100$ km.  In fact,
except for the occasional embedded moonlet (whose origin may lie in
the fragmentation of yet larger bodies rather than from the assemblage
of ordinary ring particles), the particulates in Saturn's rings have a
maximum size $\sim 5$ m \citep{zmt85}, intriguingly close to the value
implied if collective self-gravity were the only available force to
assemble the largest bodies \citep{shu84}.

Third, if we examine the most primitive meteorites, the ordinary and
carbonaceous chondrites, we find no evidence for a continuous range of
particulates spanning a range of sizes from less than a micron to,
say, a meter or more (say, comparable to the size of the entire
meteorite).  Instead, we find the largest mass fraction of such
objects to be composed of chondrules: quasi-spherical, once molten,
inclusions of millimeter and smaller sizes that give evidence of once
having been intensely and briefly heated (perhaps multiple times).
This fact suggests that sticky agglomerative events, such as those
experienced by compound chondrules \citep{rub00}, required special
transient heating events to overcome the obstacle that liquids are not
stable thermodynamic phases for any temperature at the low pressures
prevalent in protoplanetary disks.  In any case, from the meteoritic
record, the early solar system failed to generate by primitive
processes any compact particulates in excess of a few centimeters in
size (the largest refractory inclusions).  (See \citet{sea01} for a
promising, although unconventional, mechanism for producing the
refractory inclusions and chondrules in chondritic meteorites.)
Electrostatic attraction could have played a role in building looser
aggregates if the individual particulates acquired significant levels
of charge \citep{mc01}.  This effect has been seen in zero-gravity
experiments with particle densities well above the threshold for
gravitational instability.  In order to be a relevant growth
mechanism, it must be shown that tribocharging, the balance between
collisional charging and ion/electron discharging, yields
electrostatic attraction at much lower particle densities.

Fourth, even if a mechanism could be found to
grow chondrule-sized particulates to meter-sized
bodies, one would have to worry about the rapid inward orbital
drift associated with gas drag that would
carry such bodies from 1 AU into the protosun on a time scale
of only $\sim 10^2$ yr \citep{wei77}.  In contrast, mm- and km-sized bodies
have gas-drag drift times in excess of $10^5$ yr.
Only by the direct assemblage of chondrules and related objects
into planetesimals, avoiding intermediate steps,
can one prevent a rapid loss of solid material from the solar
nebula by gas drag.

Such a direct-assemblage mechanism exists in the gravitational
instability (GI) proposal put forth by \citet{gw73}.  A similar theory
was advanced independently by \citet{saf69}.  In the Goldreich-Ward
theory, particulate settling yields a subdisk of solids that is thin
and non-dispersive enough to make overdense regions undergo runaway
local contraction.  This occurs when Toomre's criterion for
axisymmetric GI in a rotating disk (the nonaxisymmetric case is
similar) is satisfied:
\begin{equation}
Q_{\rm p} \equiv {\Omega c_{\rm p} \over \pi G \Sigma_{\rm p}} < 1,
\label{eq:Q}
\end{equation}
where $\Omega$ is the angular Keplerian rotation rate, and $c$ and $\Sigma$
are the velocity dispersion and surface density.  Throughout we use
``p'' and ``g'' subscripts to refer, respectively, to the particle and
gas components of the disk.  The Goldreich-Ward instability
should not be confused
with the mechanism of \citet{bo00}, who considers the formation of
coreless gas giant planets from GI of gas disks.

Toomre's criterion (\ref{eq:Q}) is equivalent (within factors of order
unity) to the ``Roche'' limit which has been derived specifically
for the case of stratified fluids\citep{gl65,sek83}.  Sekiya finds
GI to occur when the particulate plus gas space-density, $\rho = \rho_{\rm g} +
\rho_{\rm p}$, at a distance $r$ from a star of mass $M_\ast$
exceeds a certain critical value in the midplane:
\begin{equation}
\rho > \rho_{\rm R} \equiv
0.62 M_\star/r^3. \label{Roche}
\end{equation}
At a distance $r=1$ AU from the Sun, $\rho_{\rm R} = 4\times 10^{-7}$
g/cm$^3$, which implies a critical space density of rock that
is roughly seven orders of magnitude less than the internal
density of compact rock.  Thus, in the stages preceding actual
planetesimal formation, we may treat the collection of solids
as an additional ideal ``gas'' co-mixed with the real gas of
the system.

Operating at a radius of $r=1$ AU, the self-gravitating disturbance
with the most unstable wavelength creates $\sim 5$ km planetesimals in
$\sim 10^3$ yr \citep{ys02}.  The process occurs on a time scale
longer than orbital periods ($\sim$ 1 yr in the zone for terrestrial
planet formation) because of the need to damp spin-up and random
velocities during the contraction to planetesimal densities.  But the
important point remains, that by leapfrogging intermediate size
regimes, GI avoids the rapid inspiral of meter-sized bodies.

Unfortunately, a powerful argument has been developed
against the GI scenario, which has led largely to its abandonment by
modern workers in the field \citep{wei95}.  A review of the
difficulty is necessary before before we can justify a renewed
attack on the basic idea.
 
Even in an otherwise quiescent disk, midplane turbulence may develop
to stir the particulate layer too vigorously to allow sufficient solid
settling to the midplane.  Without such settling the criterion
(\ref{Roche}) cannot be satisfied.  The problem lies in the vertical
shear possessed by disks with a highly stratified vertical
distribution of solid to gas.  The particulate-dominated sub-disk,
which possesses near-Keplerian rotation, revolves somewhat faster than
the surrounding gas disk, which has non-vanishing support against the
inward pull of the Sun from gas pressure in addition to centrifugal
effects.  The magnitude of the resulting velocity differential,
$\Delta v_\phi = \eta v_{\rm K} = \eta r \Omega$, is proportional to
$\eta$, which roughly equals the ratio of thermal to kinetic energy of
the gas:
\begin{equation}
\label{eta}\eta \equiv -{(\partial P / \partial r)\over (2 \rho_{\rm g}
r \Omega^2)} \sim (c_{\rm g}/v_{\rm K})^2,
\end{equation} 
where $P$ is the gas pressure and $c_{\rm g}$ is the isothermal sound
speed.

In the popular model of the minimum solar nebula (hereafter MSN, see
\S\,\ref{sec:prop}), $\eta \simeq 2 \times 10^{-3}
(r/\mathrm{AU})^{1/2}$, and $\Delta v_\phi \simeq 50
\,\mathrm{m/s}$ at $r=1$ AU  (Hayashi 1981).   Turbulent eddies
with a characteristic velocity equal to the available velocity differential,
$\sim\Delta v_\phi$, would then prevent GI, since the Toomre
$Q$ criterion requires
that the particle random velocity be much smaller: $c_{\rm
p} < 7 \cm/{\mathrm s} \ll \Delta v_\phi$, for instability
(Weidenschilling 1995).

These general arguments are supported by numerical simulations
\citep{cdc93} which calculate the steady state properties of two-phase
(gas and particulate) turbulence in the midplane of a MSN disk.  Particulates
with internal densities of normal rock and with
sizes from $10$ to $60 \cm$ (which are assumed to have grown by
other mechanisms and are moderately coupled to gas motions)
acquire space densities too low for GI by an order of
magnitude or more.  Their computational methods use a mixing
length prescription to relate diffusivity and
velocity shear through an extrapolation 
from laboratory studies of boundary layers.  While
quite sophisticated, this approach does not directly
address the mechanism by which the existence
of the vertical shear generates turbulence.

Using a linear stability analysis,
\citet{sek98} confirmed that GI in a turbulent dust layer is impossible,
unless the ratio of dust to gas {\it surface} densities is significantly
enhanced over normal cosmic values.  Part of the purpose of the current
paper is to understand the physical basis of Sekiya's
conclusion.  But for the present, let us merely
note that there exist several possibilities for enhancing
the solid/gas ratio in protoplanetary disks, either globally or by local
concentration.

First, in the X-wind model (Shu, Shang, \& Lee 1996), chondrules are
created from material at the magnetically truncated inner edge of the
disk, called the X-point.  Solids and gas are launched from the
X-point in a bipolar outflow.  While the gas escapes in a collimated
jet, solids of roughly millimeter size fall back to the disk at planetary
distances, thus increasing the disk's solid/gas ratio.  Since $1/3$ of all
material which passes through the X-point is launched in an outflow,
while the remaining $2/3$ accretes onto the protostar,
rocky material in the disk could be augmented by a total
amount as much as $4 \times
10^{30}~{\rm g}$, or 30 times the amount of rock in a standard
MSN model.  Such an enhancement factor is
more than sufficient, as we shall see, to promote
GI in the sub-disk of solids.   In point of fact,
because of efficiency considerations
in the manufacture of chondrules and refractory inclusions
and their irradiation to produce short-lived
radionuclides (see, e.g., Gounelle et al. 2001),
there are reasons to believe that an amount of rock not much
larger (but perhaps a few times larger) than that contained
in a MSN was recycled by the X-wind to the disk.  We shall find
that what is important is the ratio of solid to gas {\it surface
densities} in the disk.  Once we accept that this ratio need
not be cosmic (e.g., rock to gas = $4\times 10^{-3}$), then
there exists no a priori theoretical objection to a revival of the
Goldreich-Ward mechanism for forming planetesimals.

Second, since solids tend to settle towards the midplane, the surface
layers of a disk should become relatively gas-rich.  Thus, any
mechanism that removes material from the surface of a stratified disk
would increase the solid to gas ratio computed in terms of vertically
projected column densities.  Possibilities for such surface removal
include (1) photoevaporation, which dominates in the loosely bound
outer disk \citep{sjh93}; (2) layered accretion, which occurs if only
the surface layers of a disk are sufficiently ionized to support
magneto-rotational turbulence \citep{gam96}; and (3) stripping by
stellar winds, which is probably more effective near than far from the
star \citep{hyj00}.  The amounts of solid to gas enhancements
achievable by these processes are difficult to predict, but the
timescales for the dominant processes are typically $< {\rm few}
\times 10^6$ years, and thus likely to be relevant to the evolution of
protoplanetary disks.

Third, gas drag can also lead to local enhancement of particulate
concentrations by a variety of mechanisms.  (1) Isotropic turbulence
with a Kolmogorov spectrum concentrates particles in numerical
\citep{se91} and laboratory \citep{fke94} experiments.  Extrapolation
to the high Reynolds numbers of protoplanetary disks implies
concentration of chondrules by factors of up to $10^5$ \citep{cuz01}.
However, these estimates do not take into account the redispersal of
the concentrated pockets of solids if the turbulent eddies are
intermittent and do not maintain fixed centers.  (2) Similarly, disk
vortices could concentrate chondrules as well, but they are more
effective for meter sized bodies \citep{fmb01}.  There also remains
the issue whether vortices will rise spontaneously in protoplanetary
disks if there are no natural stirring mechanisms.  (3) Secular
instabilities associated with gas drag might concentrate particulates,
even without self-gravity \citep{gp00}.  Goodman and Pindor assumed
that gas drag acts collectively on a particulate ``sheet'', a valid
approximation when turbulent wakes overlap.  Unfortunately, wake
overlap seems unlikely unless the particles are fairly closely packed,
in which case GI would already be effective.  In a related context,
\citet{war76} has shown in an underappreciated study that viscous drag
modifies the standard GI criterion through the introduction of an
additional instability that can occur at values of $Q \gg 1$.
However, this additional instability, being secular in nature, has a
much smaller growth rate than the usual Goldreich-Ward mechanism.  (4) 
In \S\,\ref{sec:drift}, we develop the simplest concentration
mechanism for particulates: radial migration due to gas drag.

The fundamental assumptions required for the results of this paper
are the existence, at some epoch of the disk's
evolution, of (1) relatively quiescence in the midplane regions, even
though the surface layers may be undergoing active accretion
\citep{gam96}, and (2) compact solids with chondrule-like properties that
are well, but not perfectly, coupled to the gas through mutual drag.

Under these conditions, we argue (1) that vertical shear can only
induce a level of midplane turbulence that has limited ability to stir
solids, with the critical value of the surface density of solids being
given roughly by $\Sigma_{\rm p,c} \sim \eta r \rho_{\rm g}$; and (2)
that gas drag alone can lead to a global {\it radial} redistribution
of solids so that the local surface density of solids, $\Sigma_{\rm
p}$, can exceed the critical value, $\Sigma_{\rm p,c}$.  Therefore,
whether a recycling of solids occurs by the X-wind mechanism or a
radial redistribution of solids occurs by simple gas drag, we conclude
that the planet-forming zones of the primitive solar system can
achieve the requisite conditions for the formation of planetesimals on
a time scale comparable to the typical lifetime, $\sim 3\times 10^6$
yr, that has been inferred for the disks of T Tauri stars
\citep{hll01}.  However, the margin for success is not large, and it
could be that planet formation is a less universal phenomenon than it
has been widely touted to be, and that it has a much greater diversity
of outcomes (including a complete failure to form any planets) than
suspected prior to the discovery of extrasolar planets (Mayor and
Queloz 1996, Marcy and Butler 1998).  Indeed, the very fact that
planetesimal formation may involve a threshold phenomenon, namely the
existence of a nontrivial critical surface density, $\Sigma_{\rm p,c}
\sim \eta r \rho_{\rm g}$, implies that low-metallicity systems should
be much less likely to form planets than high-metallicity systems, a
correlation which seems already to be present in the empirical
literature (Gilliland et al. 2000, Laughlin 2000).

This paper is organized as follows.  In \S\,\ref{sec:prop} we present
the basic properties of our disk models.  After reviewing the
techniques of \citet{sek98} for deriving density distributions of
well-coupled particles in \S\,\ref{sec:tech}, we physically interpret,
in \S\,\ref{sec:sat}, the midplane density singularities which occur
in these profiles as evidence that midplane turbulence can only stir a
finite amount of material.  This allows us, in \S\,\ref{sec:enh} to
calculate the solid/gas enhancements required for GI in various model
disks.  We show that aerodynamic drift can concentrate particles of a
given size radially in the disk on cosmogonically interesting time
scales in \S\,\ref{sec:single}, and we generalize to distributions of
particle sizes in \S\,\ref{sec:dist}.  In \S\,\ref{sec:gotsat} we
evaluate whether this concentration mechanism provides enough
enhancement to induce GI.  Closing remarks are made in
\S\,\ref{sec:disc}.

\section{Disk Properties\label{sec:prop}}

If any accretion and its associated turbulence occurs in the disk,
we assume that they are confined to the top and bottom
surface layers of the disk
\citep{gam96}.  Because the deeper (midplane) layers of the disk
are then heated only by radiation from above or below, we can then
model the gaseous component as being in hydrostatic equilibrium
with a vertically isothermal distribution of temperature, while the
radial distributions for the temperature
and surface density are taken, for simplicity, to have
simple power-law profiles:
\begin{eqnarray}\label{param}
T &=& 280 f_T \varpi^{-q}\,{\rm K} , \\
\Sigma_g &=& 1700 f_g \varpi^{-p} \gsc.
\end{eqnarray}
The normalization occurs relative to 1 AU; i.e.,
we define a dimensionless radius, $\varpi \equiv r/\AU$, and
then $280 f_T$ and $1700 f_g$ represent
the gas temperature (in K) and surface density (in g/cm$^2$)
at 1 AU.  The power-law indices, $q$ and $p$, are
the remaining parameters that define our gas disk.  
In the MSN model advocated by \citep{hay81}, 
$f_T=f_g=1$, $p = 3/2$, and $q=1/2$; but for completeness,
we shall consider broader ranges of possible models.

We ignore details of the vertical structure of the gas disk, because
our interest lies in the midplane region, to which particles settle on
a timescale \citep{gw73}:
\begin{equation}\label{set}
t_{\mathrm{set}} \sim {\Sigma_{\rm g} \over \rho_s a \Omega} \sim 10^6
\, \left( {\mu\mathrm{m} \over a}\right)\;
\mathrm{yr}.
\end{equation}
In the above, $a$ is the particle radius and $\rho_{\rm s}$ is the
internal density of the
solid material, typically $3~ {\rm g/cm^{-3}}$ for rock and $1~ {\rm
g/cm^{-3}}$ for ice.

We shall initially consider
rock (r) and ice (i) distributions that reflect the
power-law distributions of the gas:
\begin{eqnarray}
\Sigma_{\rm r} &=& 7.1 f_{\rm r} \varpi^{-p} \gsc, \\
\Sigma_{\rm i} &=& \left\{\begin{array}{ccc}
0&\mathrm{if}&T(\varpi) > 170\,{\rm K}\\ 
23 f_{\rm i} \varpi^{-p} \gsc&\mathrm{if}&T(\varpi) < 170\,{\rm K}
\end{array}\right.,
\end{eqnarray}
where the total surface density of solids is: $\Sigma_{\rm p} =
\Sigma_{\rm r} + \Sigma_{\rm i}$.  At cosmic proportions,
$f_{\rm g}$, $f_{\rm r}$, and $f_{\rm i}$ would all be equal
to each other (and equal to 1 in the Hayashi model), but we
shall relax this restrictive assumption in what follows.

Table \ref{modelt} shows the parameters for the MSN and other models
used in this paper.  Values of the reference state at solar abundances
are shown.  The first letter of a model: H, A, or B; indicates its
temperature profile: warm (Hayashi's MSN values), cool, or cold;
respectively.  The surface densities are consistent with the MSN ($p =
3/2$, $f_{\rm r} = 1$) unless denoted otherwise, \eg appending an
``f'' to indicate a flatter $p = 1$ profile.

Our temperature and surface density profiles are chosen to lie within
the bounds set by astronomical observations, particularly the mm-wave
continuum emission from T Tauri disks \citep{ob95}, as well as with
the midplane temperatures predicted by the theory of passive, flared
disks \citep{cg97}.  These considerations yield icelines considerably
interior to Jupiter's 5.2 AU orbit.  Uncertainties in opacities mean
that while the particle mass in most T Tauri disks is at least that of
the MSN, it could be larger if mass is hidden in particulates much
larger than the wavelength of the observations.  While ISO detected
warm H$_2$ around T Tauri stars \citep{thi00}, more detections with
higher singal-to-noise are needed to give stringent constraints on gas
mass.

\section{The Connection Between Solid/Gas Ratio and Gravitational Instability  \label{sec:ZGI}}
 
\subsection{Dust Density Profiles \label{sec:tech}}
We briefly review Sekiya's (1998) technique for deriving dust density
profiles before interpreting the singular cusps which appear in these
profiles.  For small particles which are well-coupled to gas motions,
\ie when the particle stopping time is shorter than the dynamical and
eddy turnover times, the gas-solid mixture can be thought of as a
single stratified fluid.  This limit applies very well to mm-sized and
smaller particles, however the particles must be large enough so that
their settling times, see (\ref{set}), are shorter than the time since
global turbulence becomes weak enough to allow particulate settling.  The
well-coupled limit is the most conservative assumption
in which to demonstrate GI
since larger, decoupled solids are stirred less efficiently by gas
turbulence.

Radial hydrostatic balance yields an orbital velocity of the combined
fluid which depends on the vertical distribution or particles \citep{nsh86}:
\begin{equation}\label{azimuthal}
v_\phi(z) = \left[ 1-\eta{\rho_{\rm g} \over \rho_{\rm p}(z) +
\rho_{g}}\right] v_{\rm K} , \label{eq:rot}.
\end{equation} 
To derive the above, we have ignored the variation of
the gas density in the vertical direction since the particulate
subdisk is much thinner than a gas scale height.  When solids provide
most of the inertia, $\rho_{\rm p} \gg \rho_{\rm g}$, the fluid motion is
Keplerian, but when gas dominates we have the usual formula
for pressure-supported
rotation, $v_{\phi} = (1-\eta)v_{\rm K}$.

As particles settle towards the midplane, the vertical shear rate
(which promotes mixing) and the buoyancy (which stabilizes against
mixing) both increase.
This competition between destabilizing and stabilizing influences
is conventionally characterized by the Richardson number:
\begin{equation}\label{rich}
Ri = {N^2 \over (\partial v_\phi/\partial z)^2}.
\end{equation}
In the above
formulation, the Brunt-Vaisala frequency, $N$, is a 
measure of buoyancy: 
\begin{equation}\label{BVfreq}
N^2 \equiv
g_z\partial\ln\rho/\partial z,
\end{equation}
where $g_z$ is the vertical gravity.
When the $Ri$ drops below a critical value, $Ri_{\rm c}$,
typically $\approx 1/4$, we have Kelvin-Helmholtz instability
(KHI) and the onset of midplane turbulence.

At low particulate densities $\rho_{\rm p}\ll \rho_{\rm g}$,
we may approximate the vertical gravity as coming solely from
the central star, $g_z \approx \Omega^2 z$, while
differentiation of equation (\ref{azimuthal}) yields
for equation (\ref{rich}):
\begin{equation}\label{approxrich}  
Ri \propto {\rho_{\rm g} z
\over \partial \rho_{\rm p}/\partial z}.
\end{equation}
Increasing stratification is then destabilizing, since the increase in shear
outweighs the increasing buoyancy, decreasing $Ri$ until it reaches
$Ri_{\rm c}$.

In our detailed calculations, we include the self-gravity of the
combined fluid in the plane-parallel approximation:
\begin{equation}\label{grav}
g_z = \Omega^2z + 4\pi G\int_0^z[\rho_{\rm g} + \rho_{\rm p}(z')]dz'.
\end{equation}
When $\rho_{\rm p}$ becomes large because of the settling
of particulates to the midplane, the self-gravity of
sub-disk can provide a significant stabilizing influence
against KHI.  Indeed, we shall find that this additional effect
is the cause of a sudden ``precipitation phenomenon''
when the surface density of solids crosses a critical
value. Since the self-gravity of the subdisk of solids
is also of crucial importance for the Goldreich-Ward
mechanism, we find that ``precipitation''
is virtually synonymous with the satisfaction of
the criterion for GI, equation (\ref{Roche}).

To see this, note that for the second term on the right-hand
side of equation (\ref{grav})
to become of comparable importance as the first, we require
\begin{equation}
4\pi (\rho_{\rm g}+\rho_{\rm p}) \sim \Omega^2/G = M_\star/r^3,
\end{equation}
which differs from equation (\ref{Roche}) only by a factor
of 7.8.  Precipitation is formally reached before GI (if we
think of a process of slowly increasing the local solid/gas ratio
in the disk), but since precipitation results, as we shall see,
in the settling out of a very dense midplane-layer
of solids, GI follows almost immediately after conditions
in the disk become ripe for the beginning of precipitation.

Before these dramatic events occur,
the particulate layer should reach an equilibrium which is marginally
unstable to the KHI, because even weak turbulence is sufficient to halt the
settling of well-coupled particles.  We henceforth
assume with Sekiya that the subdisk of solids satisfies
at each value of $z$, the marginally stable condition, $Ri = Ri_{\rm c} =
1/4$, consistent with the necessary criterion for instability in
plane-parallel flows.  The result is probably modified by the disk geometry
and Coriolis forces.  However, preliminary investigations of an
analogous two-layer model (Youdin, unpublished) indicate that
instability is not significantly altered by the inclusion of rotation,
though the analog of $Ri_{\rm c}$ decreases slightly.  In any event,
meaningful quantities (particle scale height and critical surface
density) scale as $\sqrt{Ri_{\rm c}}$, so the exact
value used for the critical Richardson number
does not have a strong effect on the numerical results.

The ansatz of a marginally stable state, $Ri = 1/4$, defines via
(\ref{rich}) and (\ref{grav}) an integro-differential equation for
particle density profiles, $\rho_{\rm p}(z)$, at a given radius.  The
solution [see eqn. (18) in \citet{sek98}], requires two boundary
conditions: the midplane particulate density, $\rho_{\rm p}(0)$, and
reflection symmetry across the midplane, $\partial\rho_{\rm p}/\partial z =
0$.  The former is equivalent, as a one-to-one mapping, to an integral
constraint on $\Sigma_{\rm p}$.

Density profiles of this type can be analyzed for GI, \eg using
(\ref{Roche}).  One typically finds that low mass disks are
stable to the Goldreich-Ward mechanism at cosmic abundances
for solids/gas.  However increasing the
surface density in particles, $\Sigma_{\rm p}$, while holding
$\Sigma_{\rm g}$ fixed, leads to the development of a density cusp (see
Fig. \ref{fig:prof1}), which quickly leads to GI in the midplane,
as we described above.  Decreasing $\Sigma_{\rm g}$ relative to
$\Sigma_{\rm p}$ has the same effect.  In this context,
it is well to recall that the evidence
of the planets of the solar system tells us only what the
minimum content of the {\it solids} was in the primitive
solar nebula; it says almost nothing about either the
{\it maximum} content of solids or {\it actual} content of gas
at the time of the formation of planets and planetesimals.

\citet{sek98} was the first to find the appearance of density
cusps in the above type of analysis, but he curiously
dismisses the ``seeming infinite density''
as ``due to [an] oversimplified analysis.''  In contrast,
we believe, as already hinted above, that the appearance
of density cusps, i.e., the phenomenon of ``gravitational
precipitation'' is intimately tied to the process
of GI, and is indeed crucial to understanding how the formation
of planetesimals might occur in nature.

\subsection{Saturating the Particulate Layer -- Qualitative Understanding \label{sec:sat}}

The cusps which appear in constant $Ri$ particle density profiles
become singular, \ie reach infinite midplane volume-densities, for finite
particulate surface-densities.  Critical profiles cannot be
constructed for higher
values of $\Sigma_{\rm p}$.  We argue that the particulate layer becomes
``saturated'' at this critical surface density and excess solids
will precipitate to the midplane and undergo GI.

The value of this critical surface density is found by taking the
$\rho_{\rm p}(0)\rightarrow\infty$ limit of the general result for the
surface density as a function of midplane density, equation (22) in
\citet{sek98}:
\begin{equation}\label{anal}
\Sigma_{\rm p,c} = 2\sqrt{Ri_{\rm c}}\,\eta r \rho_g\cdot s(\psi),
\end{equation}
where
\begin{equation}
 s(\psi) \equiv (1+\psi) \ln[(1+\psi + \sqrt{1+2\psi})/\psi] -
\sqrt{1+2\psi},
\end{equation}
is an order unity term which depends only on the (typically weak)
self-gravity of the gas: $\psi \equiv 4\pi G\rho_{\rm g}/\Omega_{\rm
K}^2 \approx 1.9 / Q_g$.  The height of this critical layer can be
found similarly from equation (21) of \citet{sek98}:
\begin{eqnarray}
 H_{\rm p,c} &=&  \sqrt{Ri_{\rm c}}\,\eta r \cdot h(\psi),\\
h(\psi) &\equiv& \sqrt{1+2\psi} - \psi \ln [(1+\psi +
\sqrt{1+2\psi})/\psi].
\end{eqnarray}
These self gravitational terms can be approximated over a wide range
of $Q_{\rm g}$ values by: $s(Q_{\rm g}) \approx 1.8(Q_{\rm
g}/10)^{0.35}$ and $h(Q_{\rm g}) \approx 0.66(Q_{\rm
g}/10)^{0.18}$.

When the self-gravitational terms are neglected, we see that the
particulate mass $\Sigma_{\rm p,c} \approx \rho_{\rm g} H_{\rm p,c}$
that can be stirred by midplane shear is equal to the mass of gas in
the layer.  This result has a simple intuitive interpretation via
a dust-storm analogy.  A fierce wind-storm
in the desert can pick up a lot of dust via the Kelvin-Helmholtz
instabilities that ruffle the interface between the air and the
desert floor.  However, the wind cannot pick up the sand of the
whole desert.  There is a maximum amount of dust with which air can be
laden before as much dust falls out of the air as is picked up
by it.  In the case of a desert storm, this amount depends on
hard the wind is blowing.  In the nebular disk, the strength
of the ``wind'' is fixed by relative mechanical balance considerations
(depending on the parameter $\eta$), and
it should not be too surprising, given the lack of
intrinsic scales in the problem (when we ignore a role for self-gravity),
that this saturation level is roughly reached when the mass of dust
is equal to the mass of the layer of air in which the dust is embedded.

To derive this result more quantitatively (but still roughly),
we consider how the KHI depends on
$\Sigma_{\rm p}$ and the particle layer thickness, $H_{\rm p}$, for
low and high values of $\rho_{\rm p}/\rho_{\rm g}$.  We do this by
making the crude approximation $\partial \rho_{\rm p}/\partial z \sim
\rho_{\rm p}/H_{\rm p} \approx \Sigma_{\rm p}/H_{\rm p}^2$, which allows
us to express:
\begin{equation}\label{richapp}
Ri \sim {(\rho_{\rm g}H_{\rm p} + \Sigma_{\rm p})^3 \over (\eta r
\rho_{\rm g})^2 \Sigma_{\rm p}}.
\end{equation}
When $\rho_{\rm p} \ll \rho_{\rm g}$, the Richardson number scales as
$Ri \propto H_{\rm p}^3/\Sigma_{\rm p}$.  Thus, maintaining the balance
between buoyancy and shear as $\Sigma_{\rm p}$ increases requires
the particle layer to become slightly thicker (as is seen in the
detailed solutions), and will lead to a density increase: $\rho_{\rm
p} \sim \Sigma_{\rm p}/H_{\rm p} \propto \Sigma_{\rm p}^{2/3}$.

When the particle density becomes large, $\rho_{\rm p} \gg \rho_{\rm
g}$, it is no longer possible to maintain KHI since $Ri \propto
\Sigma_{\rm p}^2$.  While the buoyancy is relatively constant (as long
as we can still ignore self-gravity), the shear actually decreases
with added mass, $\partial v_\phi/\partial z \propto (\partial
\rho_{\rm p}/\partial z)\rho^{-2} \propto \Sigma_{\rm p}^{-1}$,
because the velocity contrast is diminished as more of the material
rotates at speeds closer to Keplerian.  From this simplified analysis
one finds that the maximum $\Sigma_{\rm p}$ which can be supported by
the KHI is: $\Sigma_{\rm p,c} \sim \sqrt{Ri_{\rm c}} \eta r \rho_{\rm
g}$, in agreement with the detailed result except for the
self-gravitational correction factor.

As already hinted upon,
saturation does not lead to GI of the entire particle layer unless the
gas disk is already self-gravitating: $Q_{\rm p} \approx \Omega^2
H_{\rm p,c}/(\pi G\Sigma_{\rm p,c}) \sim Q_{\rm g}$.  Thus,
only the unstirred particles in excess of $\Sigma_{\rm p,c}$ should
undergo GI initially.  Self-gravity is important largely for
the $s(\psi)$ factor.  
If we were to take $\psi \rightarrow 0$, which can be shown to be equivalent
to setting $g_z = \Omega^2 z$ at the outset, then as a high density cusp
develops, $\rho_{\rm p}(0) \gg \rho_{\rm g}$, one finds that $\Sigma_{\rm
p} \propto \ln(\rho_{\rm p}(0))$.  Thus, an infinite density cusp no
longer corresponds to a finite $\Sigma_{\rm p,c}$.  
Ignoring self-gravity entirely would have caused
us to miss the saturation effect in Sekiya's detailed solutions.

The fact that midplane shear can only stir a finite amount of solids
is only relevant if the saturation point can be reached.  Let us
compare $\Sigma_{\rm p,c}$ to the surface density available at cosmic
abundances :
\begin{equation}
{\Sigma_{\rm p,c} \over \Sigma_{\rm p,\odot}} \sim {\eta r \rho_{\rm g} \over 2 \Sigma_{\rm
p,\odot}} \sim  \left({\Sigma_{\rm g} \over
\Sigma_{\rm p}}\right)_\odot {c_{\rm g} \over 3v_{\rm K}},
\end{equation}
where we use $\Sigma_{\rm g} \simeq 2.4 \rho_{\rm g} r c_{\rm g} /
v_{\rm K}$ and $\eta \simeq 1.6 (c_{\rm g} / v_{\rm K})^2$.
Saturation requires that the thinness of the particulate layer,
$H_{\rm p}/H_{\rm g} \sim c_{\rm g}/v_{\rm K} \sim 1/30$, equalize the
space densities ($\rho$) of particles and gas.  Our estimate indicates
that saturation may be possible at cosmic abundances
(i.e. $\Sigma_{\rm p,c} < \Sigma_{\rm p,\odot}$ in the outer
solar system, where the inclusion of ices yields a gas/solid mass
ratio, $(\Sigma_{\rm g}/\Sigma_{\rm p})_\odot \simeq 57$.  Saturation
at cosmic abundances is unlikely if only rocks are present,
$(\Sigma_{\rm g}/\Sigma_{\rm p})_\odot \simeq 240$.  To obtain more
accurate answers, detailed solutions must be used to determine whether
and how much enhancement is required.  The simplified treatment shows
the general trends that colder disks require less enhancement.  It
also tells us that total disk mass is relatively unimportant, until the
self-gravity factor $s(\psi)$ comes into play.

\subsection{Required Enhancement Values \label{sec:enh}}

In this section we compute more accurate values for the enhancement of
solids, or the depletion of gas, required for the onset of GI in
various disk models.  For the reasons described earlier,
we adopt the ``saturation'' threshold (required to precipitate
out a midplane layer of solids with a formally infinite space density):
\begin{equation}\label{GI}
\Sigma_{\rm p} > \Sigma_{\rm p,c},
\end{equation}
as the criterion for the onset of GI.  This is clearly a more conservative
approach than considering the Roche stability criterion which only
requires a high but finite midplane space density, equation (\ref{Roche}).
In practice, because saturation is reached so quickly
once conditions become appropriate for precipitation, the two
criteria are virtually identical, except for the cases when
the gas is a considerable aid to GI.  In such extreme
conditions of massive or very cold disks, we revert to the
Toomre criterion, $Q_{\rm p}
< 1$, to assess the possibility for GI.

A disk model which is gravitationally stable at cosmic abundances can
be made unstable in several ways, as shown for the MSN in
Fig. \ref{fig:sd}.  Holding the gas content fixed, increasing
$\Sigma_{\rm p}$ will yield GI at a value, $\Sigma_{\rm p,u}$, which is
typically $\Sigma_{\rm p,c}$.  The amount of enhancement  $\mathcal{E}
\equiv \Sigma_{\rm p,u}/\Sigma_{\rm p,\odot}$ needed for GI is shown
for various models in Fig. \ref{fig:enh}.

The inverse process, holding the particle content fixed and lowering
$\Sigma_{\rm g}$ until GI occurs at a value $\Sigma_{\rm g,u}$, works
equally well.  The depletion factors, $\mathcal{D} = \Sigma_{\rm
g,\odot}/\Sigma_{\rm g,u}$, defined so that $\mathcal{D} \geq 1$, are plotted
in Fig. \ref{fig:dep}.

One can think of the gas depletion or solid enhancement scenarios as
reflecting the mechanism which gives rise to solid/gas ratio enhancements
and the physical conditions at the time of planetesimal formation.
There is no essential difference between the two procedures except
that enhancing solids yields a higher surface density.  The reason that
required depletion factors are larger, $\mathcal{D} > \mathcal{E}$, typically
by factors of 1.5-2, is the effect of self-gravity represented by
the $s(\psi)$ factor in $\Sigma_{\rm p,c}$.

We find that GI requires augmenting the particle to gas ratio by
factors of two to tens above cosmic, depending on the disk model and
radial location.  As expected, colder disks with less pressure support
(and thus less vertical shear) need less enhancement.  Higher mass
disks also require smaller enhancement factors, but the effect is
weaker.  We find that GI without enhancement is possible only in the
outer regions ($>10$ AU) of cool disks which are 10 to 15 times more
massive than the MSN in these regions.

Two caveats exist in the interpretation of these results.  If
only rocky materials, \eg chondrules, are enhanced and not ices,
then the fractional enhancement of solids
is actually $(1+W)\mathcal{E} - W$, where $W=3.2$ (or 0) is the cosmic
ice to rock ratio outside (or inside) the iceline.  Also, if
planetesimal formation occurs in one of the high $\Sigma_{\rm p}$
scenarios, the total amount of solids need not exceed the
MSN if the enhancement is local.

\section{Drift Induced Enhancement \label{sec:drift}}

Here we show that gas drag in a laminar disk causes global
redistribution and concentration of small solids as they inspiral.  It
is surprising that this straightforward and robust effect has not been
studied before, but the more complex case of the global evolution of
solids in turbulent disks has been studied in numerical simulations
\citep{sv96}.  In our case, the concentration of mm-sized solids
occurs on $\sim 10^6$ year timescales, implying that chondrules could
play a crucial role in triggering planetesimal formation and in the
observed disappearance of dust disks around T Tauri stars.

For particles with a radius $a < 9 \lambda/4$, where $\lambda$ is the
gas mean free path, gas drag follows Epstein's law \citep{wei77}.  For
the MSN, $9 \lambda/4 \simeq .6 \varpi^{11/4}\cm$, so particles up to
chondrule sizes can safely be treated with Epstein drag for all but
the innermost regions.  The drift speed due to Epstein drag in a
hydrostatic gas disk is:
\begin{equation}\label{vdr}
-{dr \over dt} \equiv v_{\rm dr} = 2{\rho_{\rm g} \over \rho} \eta
 t_{\rm st}\Omega^2 r \sim 3 \varpi^{3/2}\left({10 \rho_{\rm s}a \over {\rm
 g/cm}^2}\right) {{\rm AU} \over 10^6 ~{\rm yrs}},
\end{equation}
where $t_{\rm st} \equiv \rho_{\rm s}a/(\rho_{\rm g}c_{\rm g}) \ll
1/\Omega$ is the stopping time. The numerical values in
(\ref{vdr}) apply to the MSN; more generally the radial dependence
goes as $v_{\rm dr} \propto r^d$ where:
\begin{equation}\label{d}
d = p - q + 1/2,
\end{equation}
where we again specify $p$ as the surface density powerlaw of the gas disk:
$\Sigma_{\rm g} \propto r^{-p}$.   Note that $d$ depends only on gas
properties, which we assume to be time constant, and not on the evolving
surface density of the solids.

We will set the inertial factor $\rho_{\rm g}/\rho \approx 1$ in
(\ref{vdr}) because this approximation simplifies
the mathematics at no significant
cost in realism.  Since GI would occur (by the
saturation mechanism of \S\,\ref{sec:sat}) if
$\rho_{\rm d} > \rho_{\rm g}$, the procedure represents a factor
$\sim 2$ error at worst in non-critical circumstances.
We assume that any variation in drift speeds associated
with variations in $\rho_g/\rho$ is hidden by
the larger spread that occurs when we have a spectrum of particle sizes.
With this simplifying assumption, $v_{\rm dr}$ depends on the solid
density and size of the particles, but not on their surface or space
density.

\subsection{Evolution of a Single Particle Size\label{sec:single}}

An axisymmetric distribution of uniformly sized particles with surface
density $\Sigma(r,t)$ (we drop the ``p'' subscript here) evolves
according to the continuity equation:
\begin{equation}\label{ODE}
{\partial \Sigma \over \partial t} - v_{\rm dr}{\partial \Sigma
\over \partial r} =  {\Sigma \over r}{\partial
\over \partial r} (r v_{\rm dr}),
\end{equation}
subject to an initial value for $\Sigma(r,0)$.  This linear,
first order PDE, a Cauchy problem, can be solved by the method of
characteristics to yield a general solution:
\begin{equation}\label{gen}
\Sigma(r,t) = r^{-d-1}g(r_{\rm i}(r,t)),
\end{equation}
where
\begin{equation}
r_{\rm i}(r,t) = r\left[1 - (d-1){v_{\rm dr}(r)t \over r}\right]^{-{1 \over
d-1}},
\end{equation}
is the initial location of a particle which winds up at radius $r$ at
time $t$.  Note that the material at a fixed position $r$ is arriving
from increasingly distant locations $r_{\rm i}$ with time, regardless
of the sign of $d-1$.\footnote{For the special case $d = 1$, the
general solution (\ref{gen}) is still valid and $r_{\rm i} = r\exp[v_{\rm
dr}(r)t/r]$.}  The function $g(r) = r^{d+1}\Sigma(r,0)$ is determined
by the initial conditions.

Consider a surface density which initially has a power-law
profile with a cutoff at some outer radius: $\Sigma(r,0) =
\Sigma_0r^{-n}$ if $r<r_{\rm o}$ and $\Sigma(r,0) = 0$ otherwise.  The
surface density evolves as:
\begin{equation}
\Sigma(r,t) = \Sigma_0r^{-d-1} r_{\rm i}^{d+1-n}(r,t),
\end{equation}
if $r_{\rm i}(r,t) < r_{\rm o}$, i.e. if the material comes from
within the disk cutoff, and $\Sigma(r,t) = 0$ otherwise.

The concentration of disk material at a given location is:
\begin{equation}
\mathcal{C}(r,t) \equiv {\Sigma(r,t) \over \Sigma(r,0)} =
\left[ {r_{\rm i}(r,t) \over r}\right]^{3/2 - q +(p-n)},
\end{equation}
where the requirement that $r_{\rm i}(r,t) < r_{\rm o}$ still holds, or
else $\mathcal{C} = 0$.  We have expressed $d$ using (\ref{d}).  If
the solids and gas share the same initial powerlaw distribution, then
concentration depends only on the conditions that $3/2 > q$ (nearly
inevitable) and that the outer edge of the disk has not passed, \ie
$r_{\rm i}(r,t) < r_{\rm o}$.  The maximum enhancement at any location
takes the simple form: $\mathcal{C}_{\rm max} = (r_{\rm
o}/r)^{3/2-q}$.  This can reach factors of 10's or 100's for a disk
with $r_{\rm o} \sim {\rm few}\times 100$ AU, like observed T-Tauri
disks.  Note, however, that the largest concentration factors in the
inner disk might not be reached because of planetesimal formation that
occurs as the small solids are drifting in.

The time it takes to reach a certain enhancement level
$\mathcal{C}<\mathcal{C}_{\rm max}$ is:
\begin{equation}
t_\mathcal{C}(r) = {[1 - \mathcal{C}^{-{d-1\over 3(2-q)}}]r \over (d-1) v_{\rm dr}(r)}.
\end{equation}
For mm-sized and larger particles this timescale is shorter that the typical
disk lifetime of a few $10^6$ years. Thus, significant concentration can
be expected for chondrules and refractory inclusions.
Smaller particles, such as the
matrix material in chondritic meteorites, would not have enough time to be
concentrated by this mechanism.

The time evolution of uniformly sized ($a = 1$ mm) particles with an initial
surface density:
\begin{equation}\label{sige2}
\Sigma(r,0) = \Sigma_0 r^{-n}\exp [-(r/r_{\rm o})^2],
\end{equation}
chosen to smooth the abrupt cutoff, is plotted in
Figs. \ref{fig:dragH} and \ref{fig:dragAf}.  The particles are
embedded in untruncated gas disks corresponding to models H and Af,
respectively, and $n = p$ is assumed.  For larger $d$ values, which
correspond to high drift speeds in the outer disk regions,
concentration occurs on a shorter timescale.  Also, when $d>1$ the
enhancement is most prominent where $r_{\rm i}(r,t) = r_{\rm o}$, \ie
the radius $r$ to where the outer edge $r_{\rm o}$ has drifted by time
$t$.  Thus, the profile becomes flatter and eventually
inverted, resulting in ringlike structures.  However when $d<1$ the
inner regions are preferentially concentrated, steepening the surface
density profiles.  For the case $d = 1$ (not shown) the enhancement is
uniform with radius (for $r$ such that $r_{\rm i}(r,t) < r_{\rm o}$).

To summarize, the magnitude by which particles of a given size are
concentrated depends on the radial extent of the disk, since this
controls the amount of material available to pile up.  The timescale
depends on the drift rates, which are highest for gas that is warm and
low density and particles that are large and compact.

\subsection{Particle Size Distributions\label{sec:dist}}

Generalizing our results to the more realistic case of a spectrum of
particle sizes is straightforward since particles of a given size
evolve independently at their respective drift speed if we adopt
the basic premise that particle collisions do not lead to
particulate growth.  Initially we
assume a distribution of particle sizes such that the surface density
of particles with a size between $a$ and $a + da$ is
$\sigma_0(r)N(a)a^{3}da$.  The size distribution, initially
independent of disk radius, is typically assumed to be a simple
powerlaw, $N(a) \propto a^{-s}$, with upper and lower size cutoffs, $a_{\rm
max}$ and $a_{\rm min}$.  The mass distribution is tilted towards
large (small) particles if $s<4$ ($s>4$).

Particles of a given size evolve according to (\ref{gen}), and
integration over the size distribution yields the total surface density:
\begin{equation}\label{distev}
\Sigma(r,t) = \int_{a_{\rm min}}^{a_{\rm max}} r^{-d-1}g[r_{\rm
i}(r,t;a)]N(a)a^{3}da,
\end{equation}
where we must now take into account the dependence of $R_{\rm i}$ on
particle size via the drift speed.  Care must be taken in evaluating
this integral since the upper size cutoff, $a_{\rm max}$, varies with
$R$ and $t$, because of the requirement that material not come from
beyond a certain radius.  Even for the case of a disk with no sharp
edge, eqn.~(\ref{sige2}), a cutoff must be placed at some large finite
disk radius because, when $d > 1$, material formally drifts in from
infinity in a finite time.  This effect is not physical since the
formula for the drift speed must be modified when it exceeds the sound
speed, which occurs beyond $500$ AU for millimeter sized solids.
An insignificant amount of material lies at such radii, and we may eliminate
the mathematical problem by imposing a cutoff.  Our final results are
insensitive to the location of the cutoff as long as it
is imposed in the exponential tail, $r > r_0$.

The numerical integrations were performed using
a fifth order Romberg method to deal with the large derivatives
present in the kernel.
Figs. \ref{fig:distH} and \ref{fig:distAf} show the time evolution of
the enhancement $\Sigma(r,t)/\Sigma(r,0)$ of particles with a
distribution of sizes and masses that is characteristic of chondrules:
$a_{\rm min} = .01$ mm, $a_{\rm max} = 1$ mm, $s = 3$.  We again use
(\ref{sige2}) for $\Sigma(r,0)$, with $r_{\rm o} = 250$ AU, with the
gas disks following models H and Af.  The main effect of introducing a
size distribution is to reduce and broaden the amount of concentration since
particles of different sizes drift at different rates.

\subsection{Saturation Criterion\label{sec:gotsat}}

In order to determine whether the concentration due to gas drag only is
enough to yield GI according to the saturation mechanism of
\S\,\ref{sec:ZGI} we need to make choices about an ``initial'' size
distribution of particles.  If the particles are uniformly sized (a
false assumption) then enhancement factors are much larger than
required for GI.  Here we make a more reasonable assumption that
initially the disk contains solar abundances of matrix material with a
size distribution: $0.1~\mu{\rm m} < a < 0.1~{\rm mm}$ with $s = 4$,
augmented by 50\% (in mass) with chondrules having the size
distribution as above (in \S\,\ref{sec:dist}).  The time evolution of
the enhancement due to these two components in model Af at 1 AU is
shown in Fig.~\ref{fig:af1au}.  The maximum factor of 10 in
enhancement is almost exactly the factor 10 that is required for
marginal GI inside the iceline, see Fig.~\ref{fig:enh}.  The resultant
5 to 1 ratio of chondrules to matrix material is slightly higher than
the 4 to 1 ratio seen in ordinary chondrites.  On the other
hand, much of the matrix material in chondritic meteorites
may be chondrule fragments, so the empirical enhancement
may be considerably higher than represented by
the nominal chondrule:matrix ratio of 4:1.
In such a picture, planetesimal formation is triggered
by the appearance (plus recycling and concentration) of chondrules in the 
nebular disk, with the ratio of chondrule to pristine matrix
required to trigger GI larger in the inner solar system than
the outer, consistent with the trends running from ordinary
chondrites to carbonaceous chondrutes.

For model H at 1 AU,
concentration under the assumptions described in the
previous paragraph yields enhancement by only a factor
of 7, not quite enough for GI.  Note that if the introduction of chondrules
(say, by the rock recycling mechanism of the x-wind theory) occurs
not all at once in the beginning, but is delayed in time or is continuous
in operation, then the enhancement factors of mm-sized bodies
relative to the pristine matrix and gas (which drains continuously into the
central star) may become significantly larger than the
simple models illustrated above.

At larger radii where ice is present, concentration factors depend on
the size distribution of the icy material, a highly uncertain
quantity.  If the ice is not much larger than centimeter
sized, then its lower internal density will cause it to
concentrate on similar timescales as chondrules, and GI becomes possible.

For the somewhat arbitrary parameters we have chosen, we conclude that
aerodynamic drift provides a significant amount of concentration, but
not definitively enough to cause midplane GI.  Thus, in the case
of our own solar system, one of the other
solid/gas ratio enhancing mechanisms discussed in the introduction may be
necessary.

\section{Discussion \label{sec:disc}}

In this paper we have shown that planetesimals can form by midplane
gravitational instabilities despite Kelvin-Helmholtz stirring if the
ratio of particle to gas surface densities is increased above cosmic
abundances.  We have also presented a simple drift mechanism which can
provide most or all of the particle enhancement required for this to
occur.  This mechanism has the advantage that it must operate in
passive protoplanetary disks, and does not depend on assumptions about
accretion physics.  There are some attractive features of this
scenario not yet discussed.

Curiously, the mass in our planetary system appears to be
significantly truncated outside of 40 AU, \ie in the Kuiper belt
region and beyond \citep{tb01}, whereas T-Tauri disks typically extend
to several hundred AU.  The size disparity is even greater if Uranus
and Neptune migrated from a location interior to Saturn's orbit
\citep{tdl99}.  The process of drift induced enhancement offers an
explanation.  The outer disk is drained relatively of its solid
resources by inward particulate drift, and thereby becomes or remains
inhospitable to the formation of planetesimals.

Another troubling aspect of planet formation theories is the
``Type-I'' migration of Earth-sized and larger (but not large enough
to open a gap) bodies due to density waves torques exerted on the gas
disk \citep{war97}.  When the resonant torques are assumed to damp
locally in the disk, Earth-mass bodies (at 5 AU) migrate inwards in
$10^5$ years in a MSN disk.  This timescale is inversely proportional
to the mass of the body, but increases sharply once the gap-opening
mass, typically $10-100 M_\oplus$ is reached.  However the drift speed
is proportional to $\Sigma_{\rm g}$.  Thus if planetesimals form due
to depletion of gas below MSN values (or equivalently the enhancement
of solids in a very low mass disk), then subsequent earth-mass cores
suffer less migration.  Gas depletion by one order of magnitude would
significantly increase the survival odds of a nascent planetary
system, and still leave enough gas for the formation of giant planet
atmospheres.

To summarize, in conventional cosmogonies with unit sticking
probabilities and cosmic abundances of solids and gas in a MSN
\citep{lis93}, planetesimal formation is easy, occurring on a time
scale of $\sim 10^4$ yr or less, while giant planet formation is hard,
requiring time scales in excess of the typical lifetimes of T Tauri
disks, $\sim 3 \times 10^6$ yr.  Overall surface density enhancements
(gas and dust) above MSN values can speed up post-planetesimal growth 
\citep{tdl02}, but this solution would exacerbate the problem of the
Type-I migration of planetary embryos (Ward 1997).  It would also
require a finely tuned mechanism to remove the considerably greater
amount of extra gas and solids from the solar system.

In an unconventional cosmogony, where sticking probabilities are zero
(except for the special mechanisms that produce chondrules and
refractory inclusions near the protosun), where gas is depleted, and
solids are enhanced relative to standard MSN values, the conditions
for the formation of planetesimals and giant planets might be
intimately tied to the gas-dust evolution of the nebular disks of T
Tauri stars.  The two processes would then naturally acquire similar
time scales, related by a single continuous process of gravitational
growth in a gas-dust disk.  Attractive byproducts of this
unconventional approach would be a corresponding alleviation of the
problems of Type I migration and gas-disk dispersal, as well as a
possible understanding of why the Kuiper belt marks a sudden apparent
truncation of the primitive solar system.

\acknowledgments

A.~Y.~would like to thank Andrew Cumming, Jeff Cuzzi, Greg Laughlin,
Geoff Marcy, and Jonathan Swift for helpful discussions and the
referee for suggesting the following appendix and other improvements.
Financial support was provided by a grant from the NASA Origins of
Solar Systems Program.  A.~Y.~acknowledges support from an NSF
Graduate Fellowship.

\appendix
\begin{center}
{\bf APPENDIX}
\end{center}
\section{Collisional Effects}
In most of this paper we have ignored the effects of collisions.  Here
we argue that this approach is is dynamically justified.  The time
between collisions is:
\begin{equation}
t_{\rm coll} \simeq {\rho_{\rm s} a H_{\rm p} \over \Sigma_{\rm p} c_{\rm
p}} \approx {\rho_{\rm s} a \over \Sigma_{\rm p}\Omega},
\end{equation}
which agrees within factors of order unity with the more exact results
presented in \citet{ws93} if we use $c_{\rm p}/H_{\rm p} \approx
\Omega$ in the second equality.  Thus the collisional timescale is
longer than the stopping time, $t_{\rm coll}/t_{\rm st} \approx
\Sigma_{\rm g}/\Sigma_{\rm p}$ by at least an order of magnitude.

Even if collisional dynamics were able to introduce an effective
viscosity, $\nu_{\rm coll} \simeq c_{\rm p}^2t_{\rm c}$ \citep{gt78}, the
characteristic diffusion velocity, $v_{\rm coll} \simeq \nu_{\rm coll}/r$
is less than the drift velocity due to Epstein drag:
\begin{equation}
{v_{\rm coll} \over v_{\rm dr}} \simeq \eta{\Sigma_{\rm
g} \over \Sigma_{\rm p}} \lesssim 10^{-2}.
\end{equation}

If collisions could change the size distribution of solids due to
fragmentation or mergers then our results for drift induced
enhancement (\S\,\ref{sec:drift}) would have to be modified by
including a coagulation equation, \eg \citet{kb02}.  However we
already argued in the introduction against the effectiveness of
collisional agglomeration.  As for fragmentation, the presence of
significant numbers of intact chondrules (the species most
important for the enhancement) in meteorites shows that many were able
to survive their collisional history in the nebular disk
without much shattering.

\clearpage

\plotone{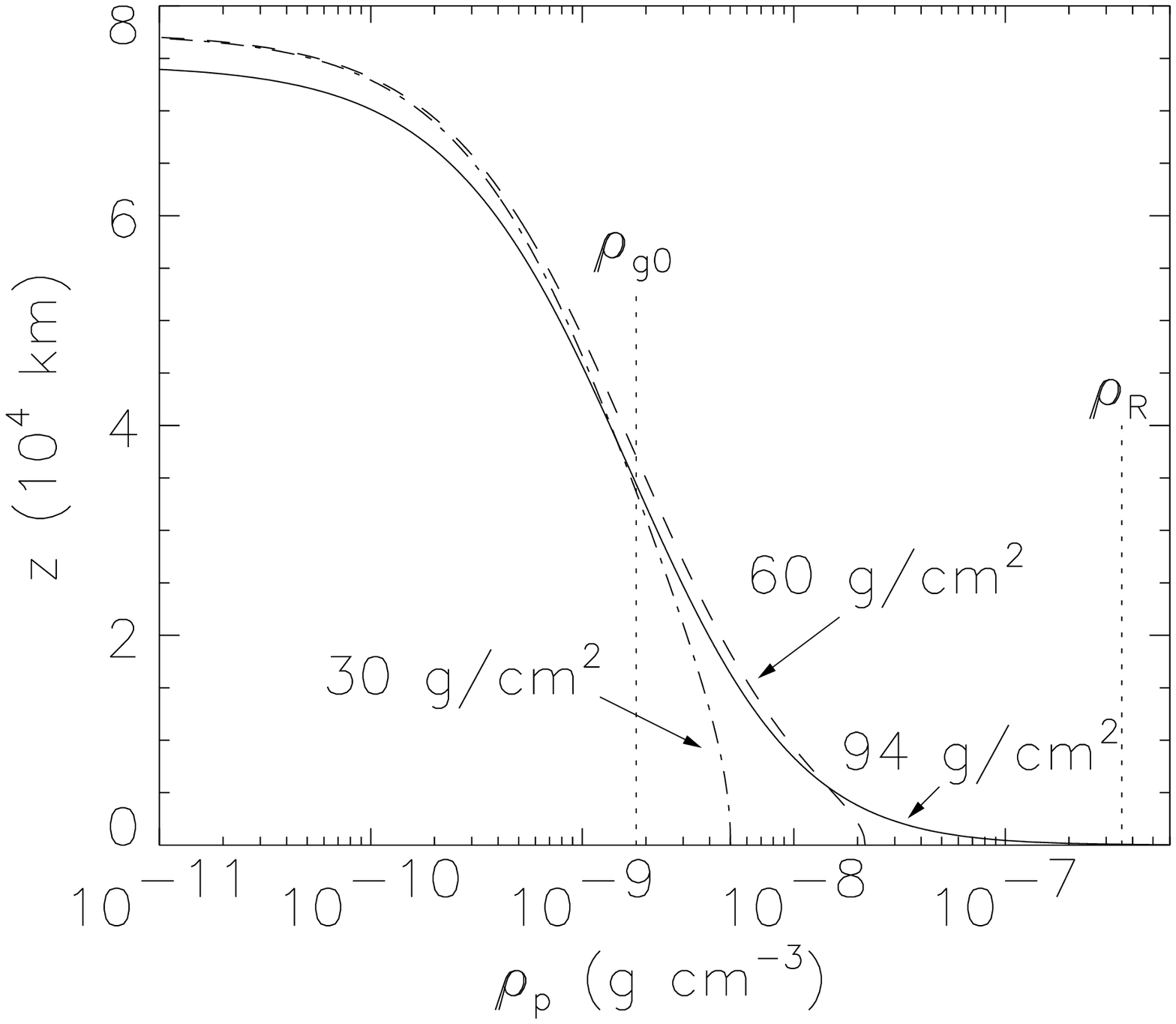} \figcaption[f1.eps]{Vertical profiles of
particle density at 1 AU, for small solids stirred by the
Kelvin-Helmholtz instability.  The gas is described by model A and
held fixed.  As $\Sigma_{\rm p}$ increases from $30 ~{\rm g/cm^2}$,
the value at solar abundances (with ice), a midplane density
cusp develops, becoming infinite at $\Sigma_{\rm p} = \Sigma_{\rm p,c}
= 94 ~{\rm g/cm^2}$ for this model.  Also shown are the gas, $\rho_g$,
and the Roche, $\rho_R$, densities. \label{fig:prof1}}

\plotone{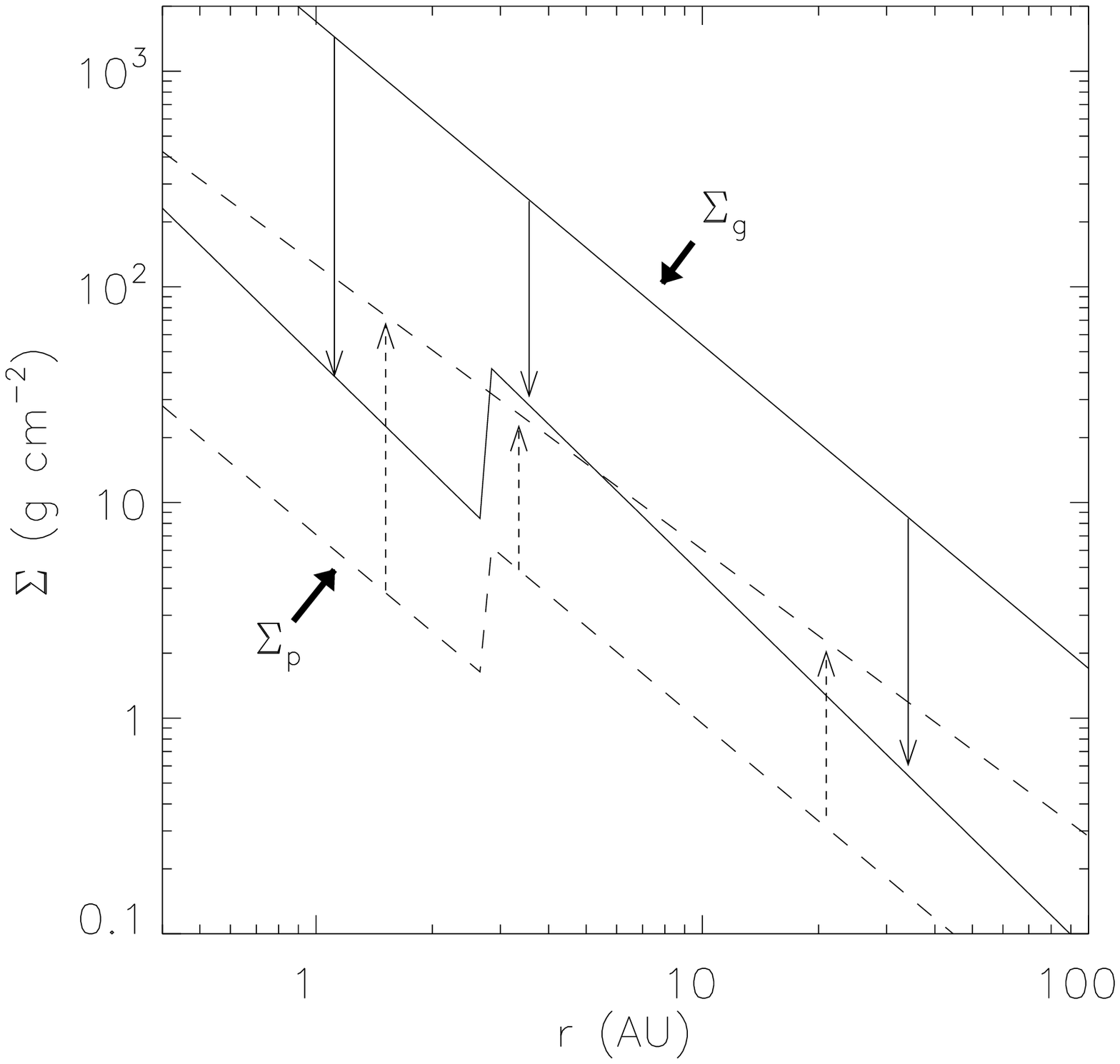}\figcaption[f2.eps]{The MSN disk model can be made
gravitationally unstable by either enhancing the solids (increasing
$\Sigma_{\rm p}$ as indicated by the dashed arrows), by depleting the
gas (decreasing $\Sigma_{\rm g}$ as indicated by the solid arrows), or
by some combination of the two (not shown).  Note that instability is
local, and need not occur simultaneously at different
radii. \label{fig:sd}}

\plotone{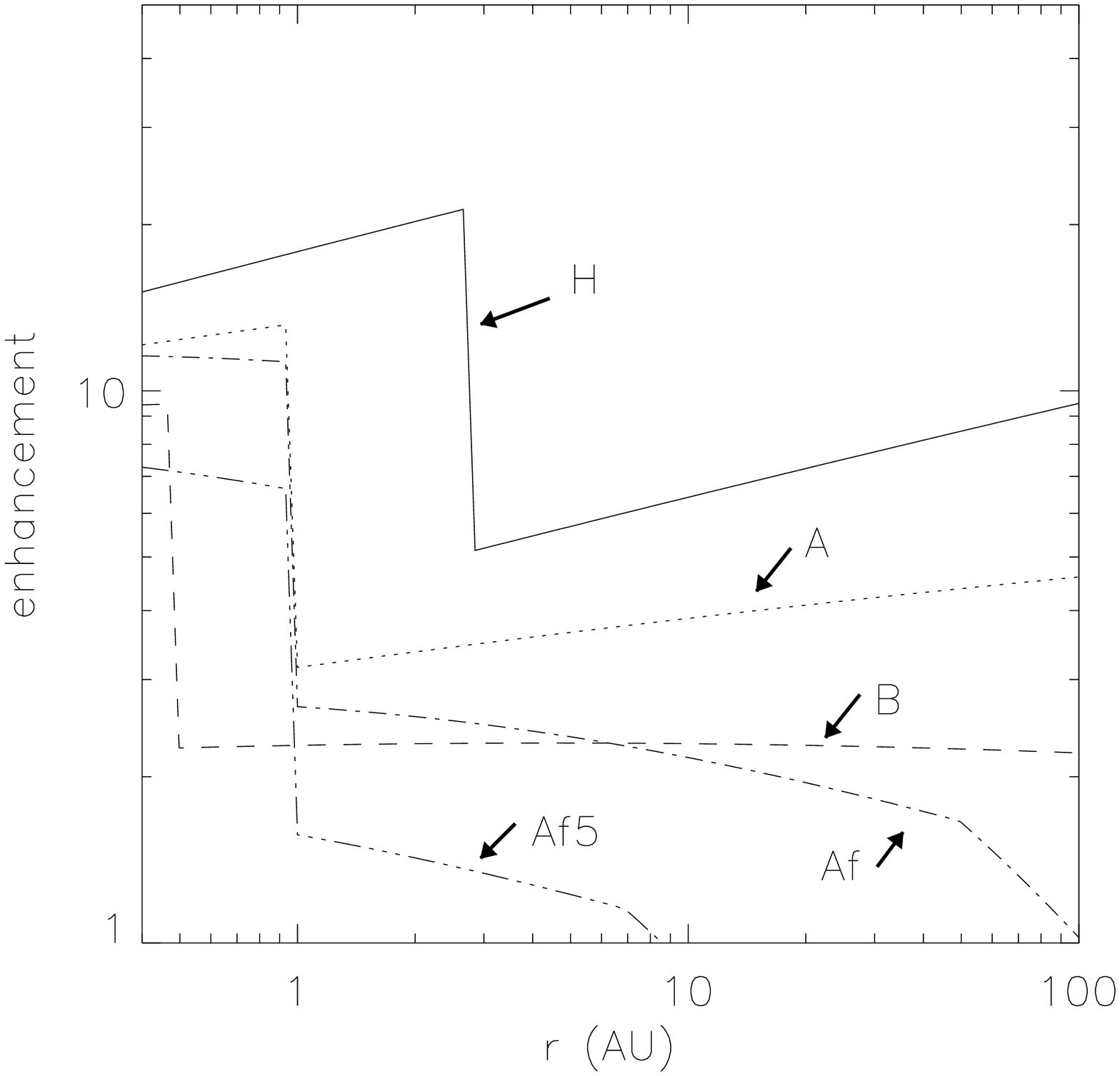} \figcaption[f3.eps]{Enhancement factor,
$\mathcal{E}$, required for GI vs. radius for various models.  The
discontinuity occurs because larger enhancements are needed inside the
iceline.  Kinks in higher mass models (Af,Af5) occur when $Q_{\rm p} <
1$ occurs before saturation.
\label{fig:enh}}

\plotone{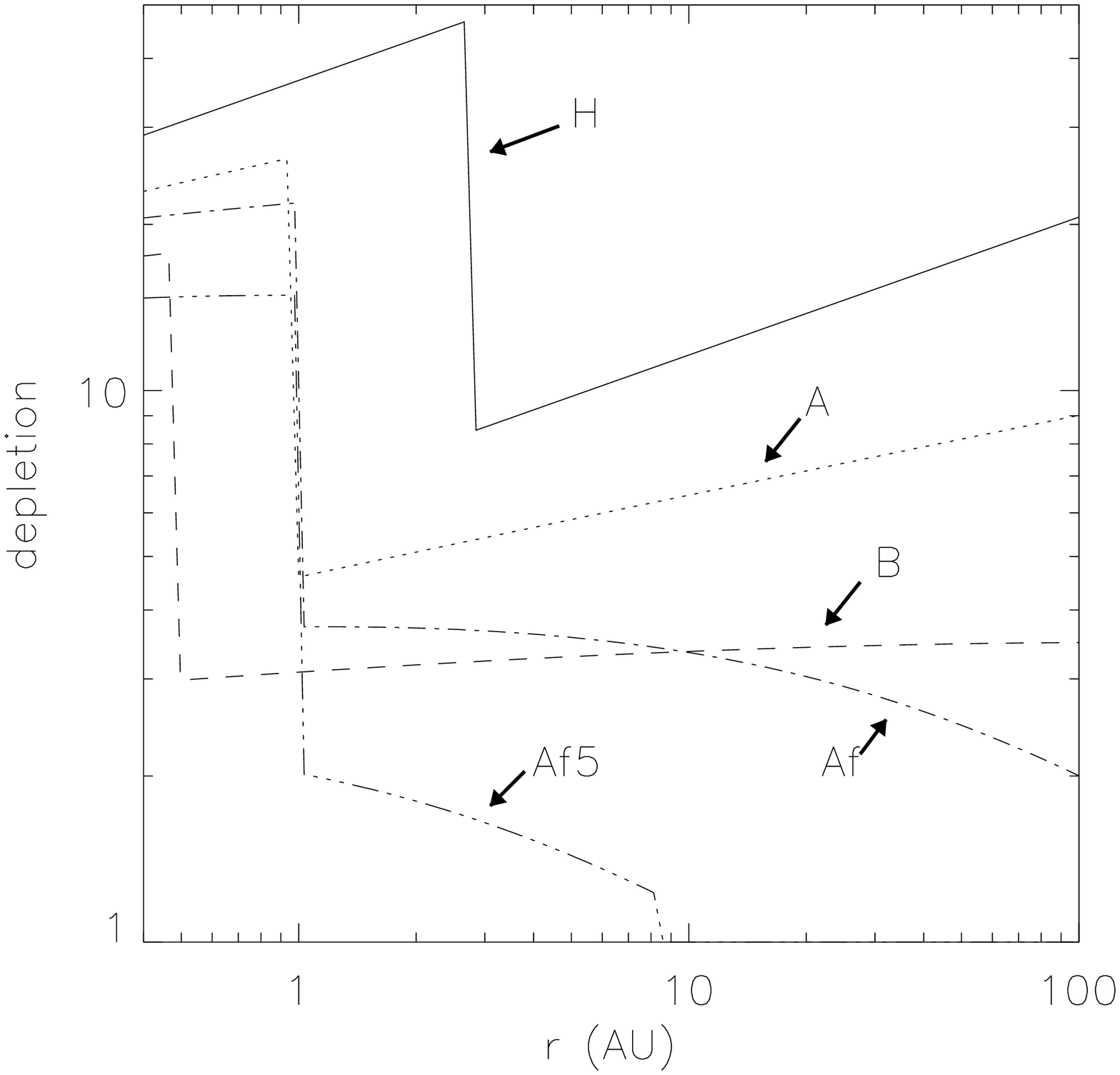}\figcaption[f4.eps]{Depletion factor,
$\mathcal{D}$, required for GI vs. radius.  \label{fig:dep}}

\plotone{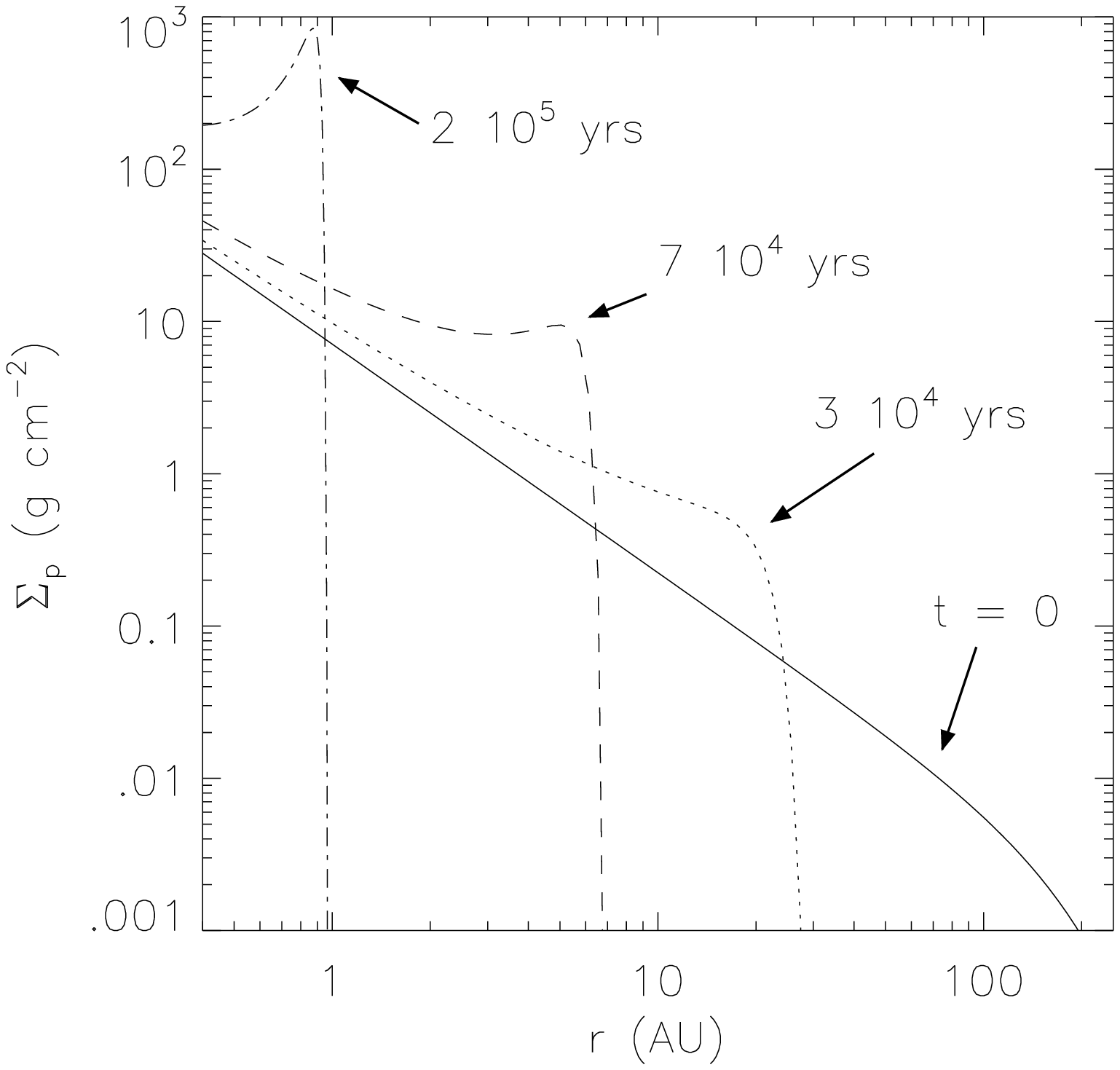}\figcaption[f5.eps]{Time evolution of the
surface density of mm-sized particles with solid densities, $\rho_{\rm
s} = 3$ g/cm$^3$, due to Epstein drag.  The gas is assumed fixed at
MSN values.  Since $d = 3/2 > 1$, the $\Sigma_{\rm p}$ profiles,
initially described by eq.(\ref{sige2}) with $R_{\rm o} = 200$ AU,
become flatter and eventually inverted into ringlike
structures. \label{fig:dragH}}

\plotone{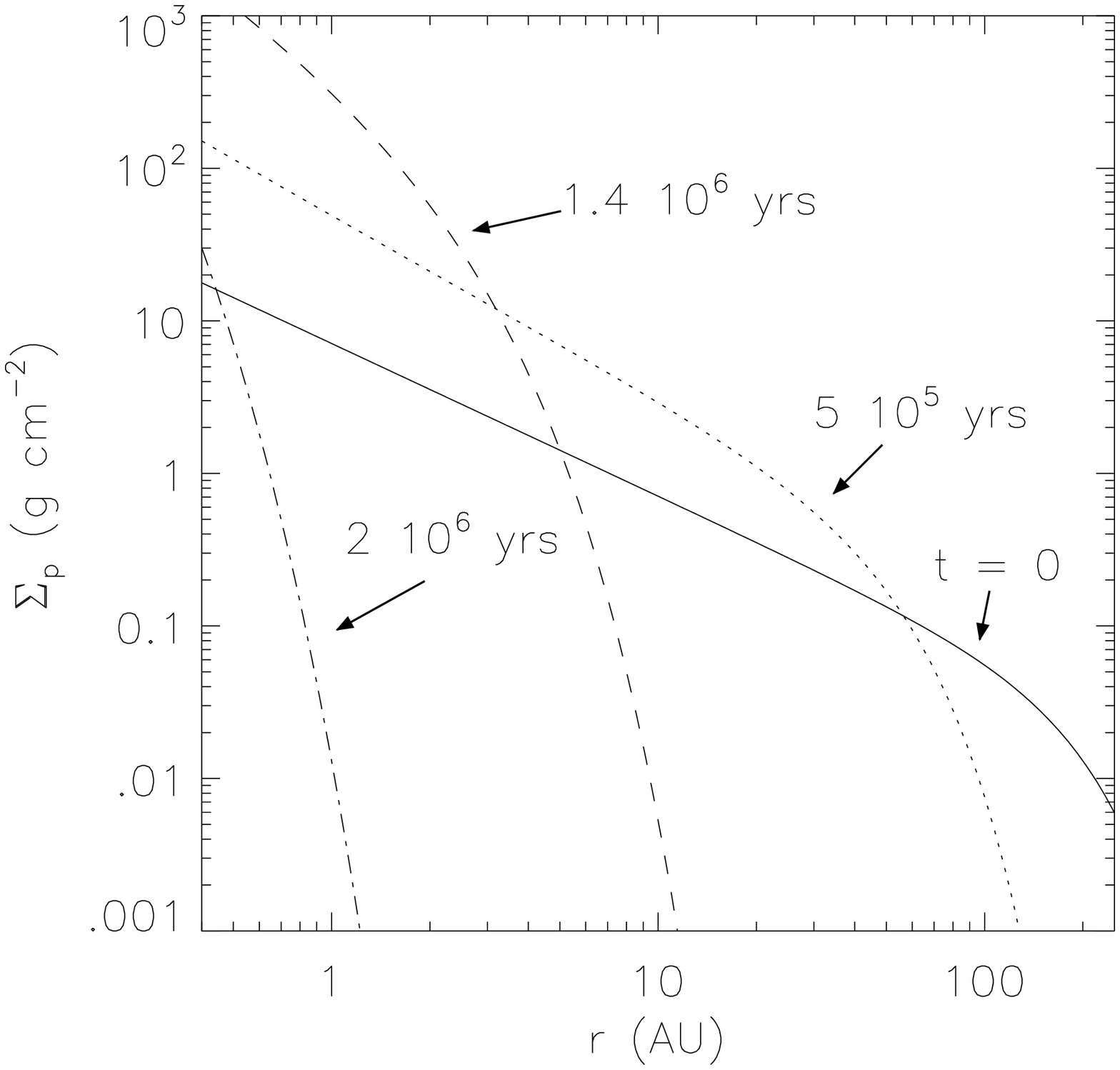}\figcaption[f6.eps]{Same as
Fig. \ref{fig:dragH} except the gas is described by model Af.  Since
$d = .87 < 1$, $\Sigma_{\rm p}$ evolves to steeper
profiles. \label{fig:dragAf}}

\plotone{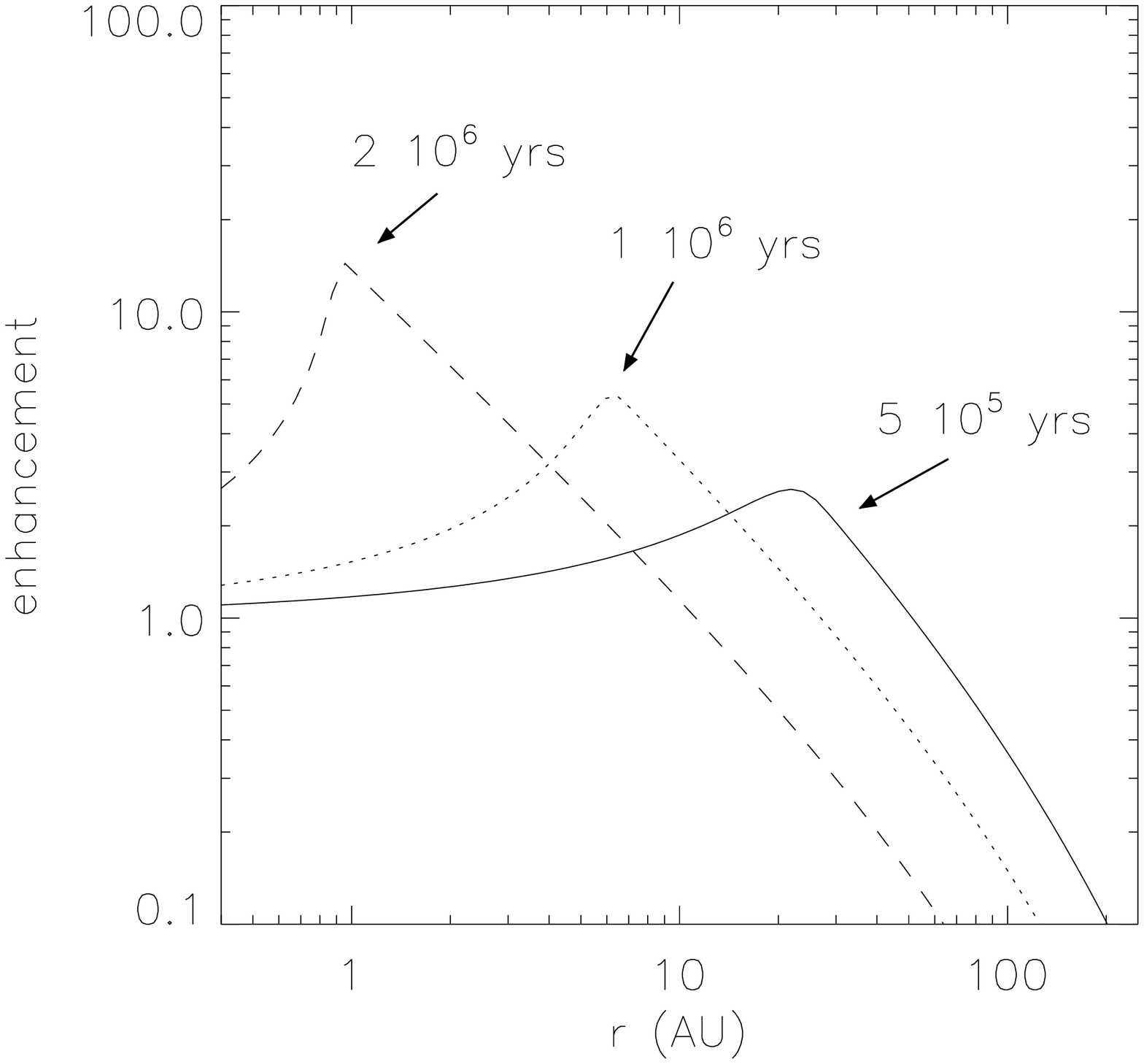}\figcaption[f7.eps]{Time evolution of the surface
density enhancement of solid particles with a size distribution
characteristic of chondrules.  The gas disk is described by the MSN.
See text for details.
\label{fig:distH}}

\plotone{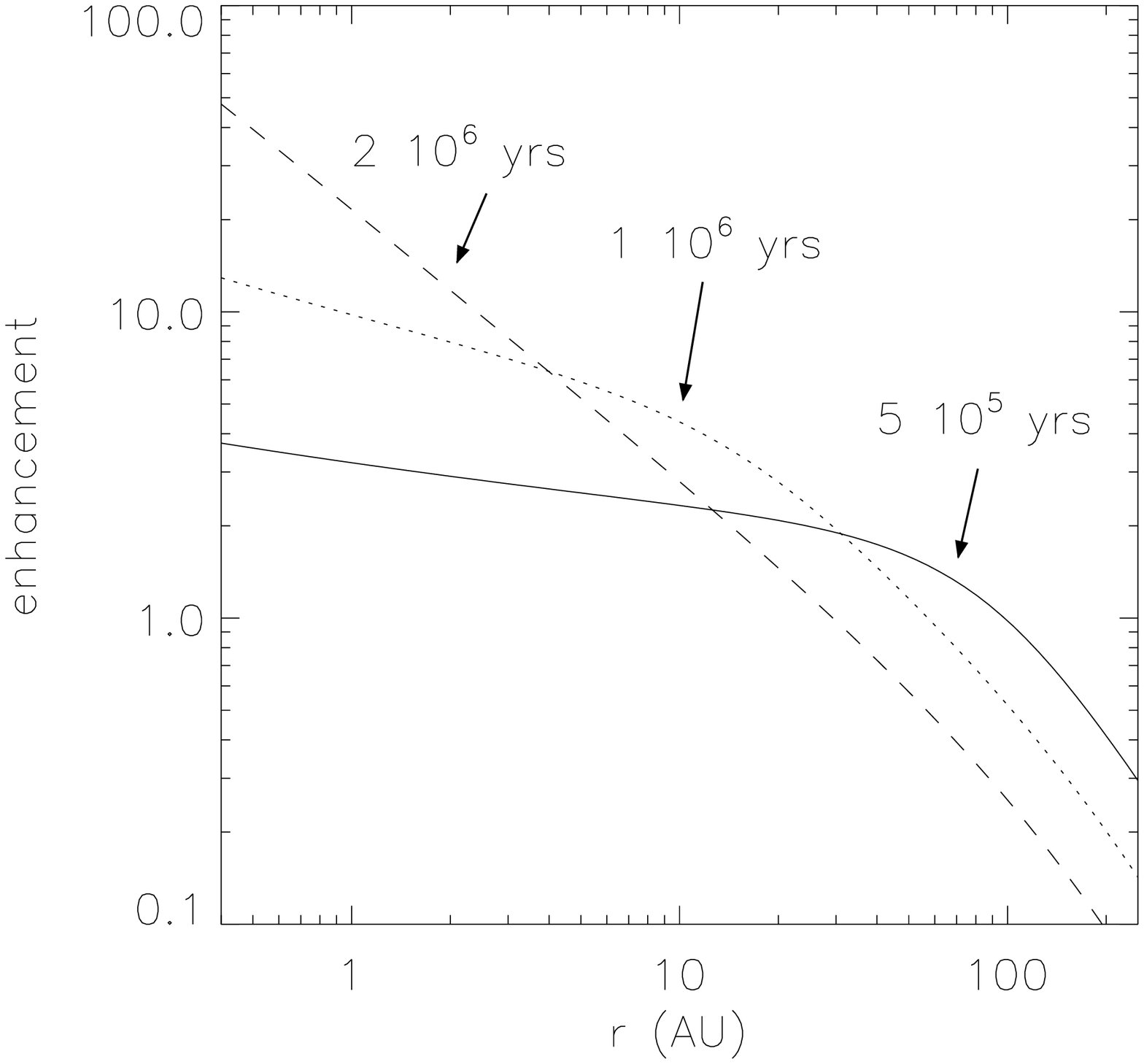}\figcaption[f8.eps]{Same as Fig. \ref{fig:distH}
except the gas is described by model Af. \label{fig:distAf}}

\plotone{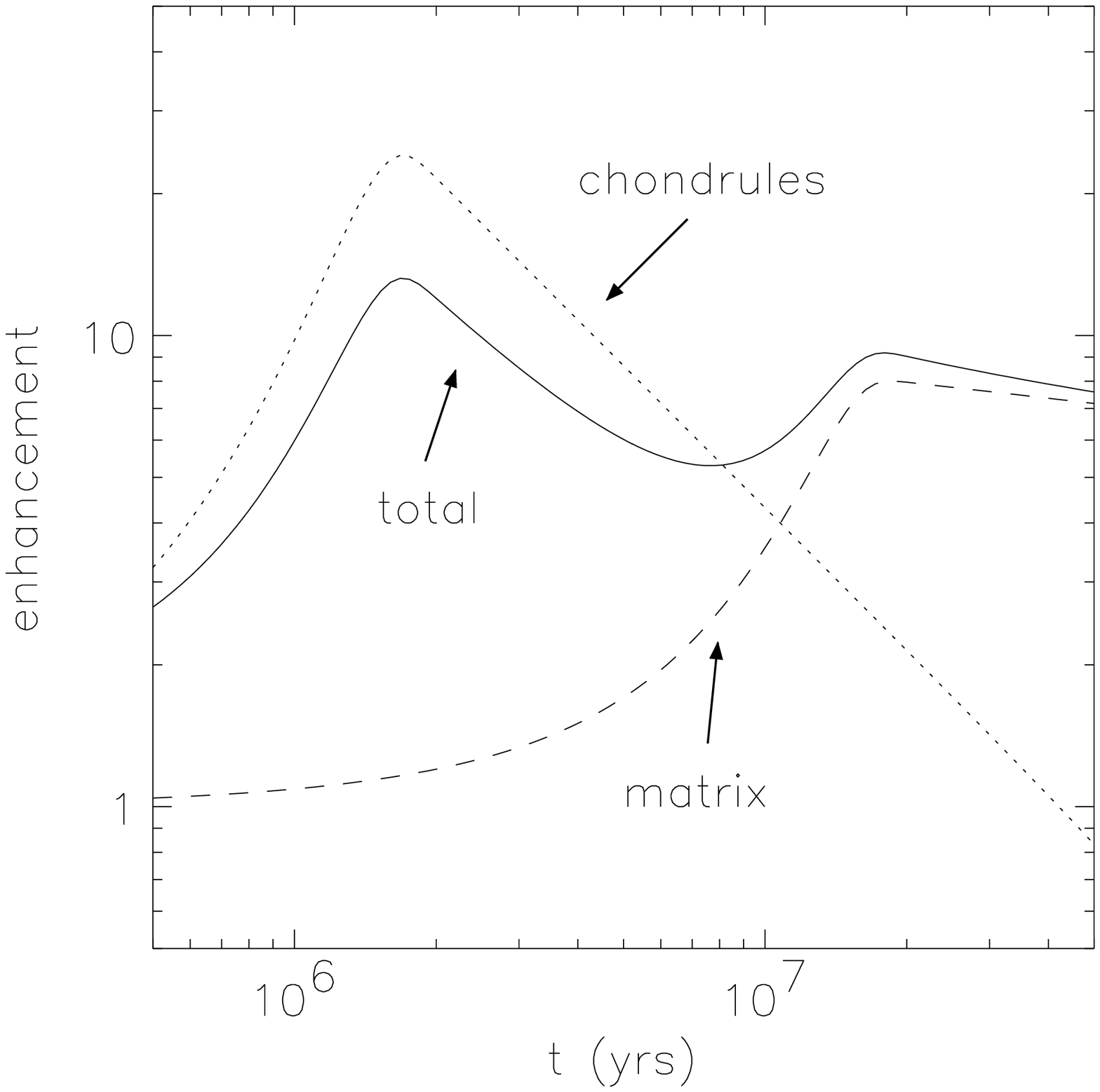}\figcaption[f9.eps]{Enhancement of solids vs. time
at 1 AU in model Af for a particle disk that is $1/3$ chondrules
and $2/3$ matrix material by mass. \label{fig:af1au}}

 \begin{deluxetable}{ccccccc}
\tablewidth{0pt} \tablecaption{Model Parameters \label{modelt}}
\tablehead{ \colhead{Model} &\colhead{$p$} &\colhead{$f$\tablenotemark{a}}&\colhead{$q$}
&\colhead{$T_1$ (K)} &\colhead{iceline (AU)}&
\colhead{$d$\tablenotemark{b}}}

\startdata
H&3/2&1&1/2&280&2.7& 1.5 \\ 
A&3/2&1&.63&170& 1& 1.37\\ 
B&3/2&1&3/4&100& .5& 1.25\\
Af&1&1&.63&170& 1& .87\\
Af5&1&5&.63&170& 1& .87\\ 
\enddata 

\tablenotetext{a}{$f = f_{\rm g} = f_{\rm i} = f_{\rm r}$ for solar
abundances.}

\tablenotetext{b}{See \S\,\ref{sec:drift}, eq. (\ref{d}).}

\end{deluxetable}
\end{document}